# Micro-gap Thermo-photo-thermionics: An Alternative Approach to Harvesting Thermo-photons and its Comparison with Thermophotovoltaics

Ehsanur Rahman[1,2]* and Alireza Nojeh[1,2]

[1] Department of Electrical and Computer Engineering, University of British Columbia, Vancouver, BC, V6T 1Z4, Canada

[2] Quantum Matter Institute, University of British Columbia, Vancouver, BC, V6T 1Z4, Canada

*Email: ehsaneee@eee.buet.ac.bd

**Abstract**

This work investigates an alternative to thermophotovoltaics for harvesting thermal and optical energy via photon coupling and thermionic energy conversion. In this device, a heat source is radiatively coupled to a thermionic electron emitter through a nanoscale gap and the electron emitter is coupled to collector through a microscale gap. The analysis using fluctuational electrodynamics and finite-time thermodynamics shows that for identical thermal radiator and photon-to-electron conversion materials, the output power density in the thermionic device can be more than twice that of the thermophotovoltaic device; the thermionic mechanism can also provide more than 30% improvement in the energy conversion efficiency compared to the thermophotovoltaic device. Moreover, the maximum-power-point voltage in the thermionic device is shown to significantly exceed the conversion material's band gap, which determines the theoretical upper limit of the open-circuit voltage in a thermophotovoltaic cell. Therefore, the results of this study indicate that thermionic energy harvesting via thermo-photon coupling can be a promising alternative to thermophotovoltaics.

# 1. Introduction

Thermal and optical energy harvesting for electrical power generation are of great interest due to the widespread availability of free heat and light (*e.g.,* geothermal and solar energy) as well as the ever-growing demand for electricity in diverse applications. Heat and light can also be converted to one another, and heat is also generated by burning fossil fuels or as a byproduct of other chemical processes, being released partly as waste to the environment during its conversion to other forms of energy [1]. Considering these factors, static heat-to-electricity conversion is of significant importance due to its various advantages over conventional generation methods relying on turbine technologies. Thermophotovoltaics (TPV) is a prime candidate in this regard [2–6], being based on conventional photovoltaics (PV), which are commercially well established.

Originally, thermophotovoltaics was motivated by the need for more efficient utilization of solar energy. The idea consists of creating a secondary thermal radiation source, which can be manipulated for spectral matching of the PV material's band gap to minimize the proportion of photons that would either cause detrimental heating (above-band gap) or go unutilized (sub-band gap). Moreover, the thermal radiator in a TPV device can be energized from a wide variety of sources, thereby enabling the TPV mechanism to harvest energy of non-solar origin. For example, TPV devices have been considered for space and terrestrial electricity generation using different sources [7–11]. Due to these prospects, various efforts have been made to improve the conversion performance of TPV devices such as through manipulation of the spectral response of the thermal radiator [12–16], introduction of light-pipes [17] or reduction of the gap size between the thermal radiator and TPV cell to enhance the radiative coupling [18–20] (also known as the near-field thermophotovoltaics (NF-TPV)).

However, harvesting thermal energy using photovoltaics inherently shares the challenges of solar energy harvesting (*e.g.*, sub-band gap photon loss and above-band gap thermal loss). An additional challenge of TPV is to maintain a high temperature difference between the thermal source and the converter. This is because thermal radiation in practically achievable temperature ranges has a large infrared portion, thereby requiring low-band gap semiconductors to harvest it. This property exacerbates the heating-related deterioration of the PV mechanism (*i.e.*, it leads to an increase in dark current and a related reduction in the open-circuit voltage compared to solar PV), thus making thermal management more demanding and possibly even requiring cryogenic solutions [21]. This is especially true for NF-TPV [22], where the photon-flux is significantly enhanced due to the coupling of evanescent waves. The latter approach has been recently experimentally shown to achieve a 40 fold enhancement in power density (compared to the far-field regime) at a gap distance of less than 100 nm [23]. However, the power density (~ 30 nW) and efficiency (~ 0.02%) reported for that device are too small for practical use. Other recently reported NF-TPV devices have similar low-performance metrics with an efficiency of less than 1% [24] and an output power density of the order of 1 $\mu$W cm$^{-2}$ [25]. A more recent NF-TPV device has been reported with a power density of 0.75 W cm$^{-2}$ and an efficiency of 14% (not considering the cooling power) [26], but the PV cell had to be maintained at a cryogenic temperature of 77K. On the other hand, higher conversion efficiencies have been reported in far-field TPV devices [27–29]. However, the power density of these far-field counterparts is less than 1W cm$^{-2}$.

Considering the above challenges of TPV, here we present an alternative approach to harvesting thermo-photons by using thermionic emission, which can outperform thermophotovoltaics. Thermionic energy conversion has a unique advantage in that it relies on thermal energy to excite electrons and, therefore, fundamentally requires high temperature [30, 31]. The thermal boost given to the electrons in the thermionic process can result in much higher open-circuit voltages compared to TPV (the theoretical maximum of which is set by the PV material's band gap).

A natural question arises as to how to interface the thermionic energy converter with the heat source. The default practice is to simply place the thermionic emitter in thermal contact with the heat source. However, in analogy with the TPV structure, one can also introduce a vacuum gap between the thermionic emitter and the heat source [32]. The premise is that this gap provides an additional degree of freedom in optimizing device performance. The energy exchange between the heat source and the thermionic emitter will now occur through thermal radiation, similarly to a TPV structure. We emphasize that the thermionic emitter will not only take advantage of the heat generated by the received radiation, but also, can experience an optical upshift of the thermal electron population, which further reduces the emission barrier for thermionic electrons, similarly to photon enhanced thermionic emission solar cells [33–37]. We show that, for identical semiconductor and thermal radiator materials, a thermo-photo-thermionic device can significantly outperform its thermophotovoltaic counterpart on several key conversion performance metrics such as power density, conversion efficiency, output voltage etc. Moreover, we show how the vacuum gap between the thermal radiator and thermionic emitter can improve the conversion performance of the thermionic device instead of maximizing the emitter temperature via direct thermal contact. Therefore, this work presents both an alternative to TPV for harvesting thermo-photons using thermionic emission and comprehensive analysis of the thermionic device's

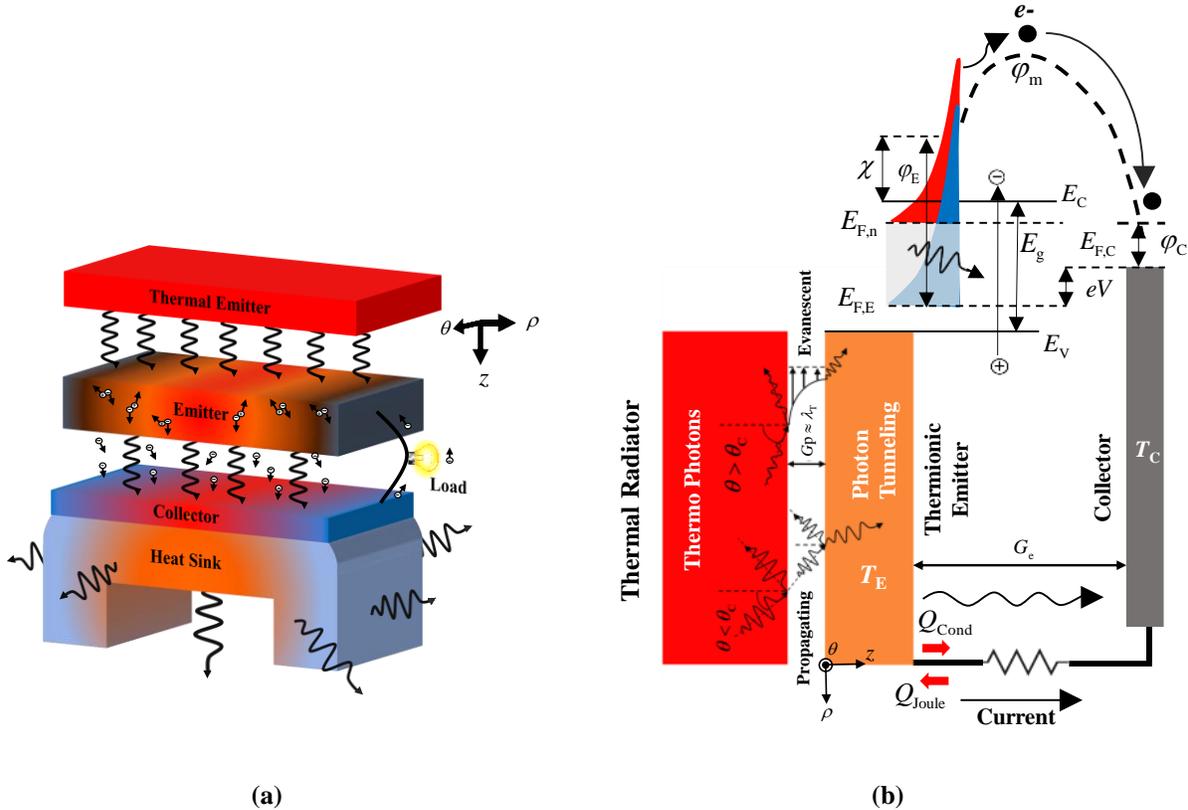

**(a)**                                                **(b)**

operation in both the near-field and far-field regimes of thermo-photon conversion.

**Fig. 1(a)** Device schematic of a TPT converter. **(b)** Operational mechanism and energy band diagram of the TPT converter. $E_{F,E}$ and $E_{F,C}$ are the equilibrium Fermi levels in the thermionic emitter and collector, respectively, and $E_{F,n}$ is the optically excited emitter quasi-Fermi level for electrons. $\varphi_m$ is the maximum motive in the interelectrode space measured with respect to the equilibrium Fermi level and $e$ is the electron charge. $\varphi_E$ and $\varphi_C$ are the work functions of the thermionic emitter and the collector, respectively, and $\chi$ is the electron affinity of the thermionic emitter. $T_E$ and $T_C$ are the temperatures of the thermionic emitter and the collector, respectively. $E_C$, $E_V$ and $E_g$ are the conduction band minimum, the valance band maximum and the band gap energy, respectively. $V$ is the voltage difference between the two electrodes. $\theta_C$ is the critical angle of incidence at the thermal radiator-vacuum interface; $G_p$ and $G_e$ are the photon and electron coupling gap sizes, respectively and $\lambda_T$ is the characteristic wavelength of thermal radiation, given by Wien's displacement law. $Q_{Cond}$ and $Q_{Joule}$ are the thermal conduction and Joule heat loss via the electrical leads of the device.

## 2. Methodology and Model Validation

### 2.1. Thermo-photo-thermionic device

#### 2.1.1. Operational Mechanism

First, we describe the operation of a thermo-photo-thermionic (TPT) converter as shown in fig.1 (a)-(b). The device consists of a thermal radiator, which converts the input energy to photons. The input energy can be from various sources such as solar radiation (which can be captured using a selective absorber or by exchanging heat from a solar thermal fuel), industrial waste heat or other chemical processes. The thermo-photons generated by the thermal radiator will be absorbed by the thermionic electron emitter (also known as the cathode) which is separated from the radiator by a vacuum gap, $G_p$ (which we term as the photon coupling gap). The above-band gap photons generate electron-hole pairs that will diffuse through the emitter material (which is p-doped to capitalize on the photon enhancement of thermionic emission [34]). The electrons that reach the emitter surface and have sufficient energy to overcome the material work function and the space charge potential barrier (see fig. 1b) emit thermionically and traverse the electron coupling gap, $G_e$, and are absorbed by the electron collector (also known as the anode). The energy carried by these thermally excited electrons is thus used to drive an external electrical load. Apart from the thermionic exchange of energy, there is thermal radiative exchange between the electrodes as well as thermal conduction and Joule heat loss in the leads. The photons which are radiated back from the thermionic emitter towards the thermal radiator are absorbed and recycled in the radiator and will help increase the conversion performance.

At first glance, it might appear that, for best performance, physical contact between the electron emitter and the source of heat is required, and that optical coupling across a gap may be detrimental. However, as our results will show, crucially, that is not the case and the inclusion of such a gap can lead to better performance. This is the premise on which the concept of TPT is based. Next, we will discuss the implementation of the various physics involved in the TPT device.

## 2.1.2. Thermionic emitter model for calculating the spatial distribution of thermo-photon absorption and photogenerated electron-hole pairs

The photons from the thermal radiator which have energies greater than the bandgap of the thermionic emitter, will be absorbed by it and generate electron-hole pairs. The spatial distribution of these photogenerated electrons in the emitter, $n(z)$, can be obtained from the particle continuity equation as [38,39]

$$D_n \frac{d^2 n}{dz^2} = R(z) - G(z) \quad (1)$$

, where $D_n$ is the electron diffusion constant and $R(z)$ and $G(z)$ are the net recombination and generation rates, respectively. The recombination rate has contributions from radiative and non-radiative (Shockley-Reed-Hall, Auger and surface) mechanisms. The generation rate depends on the photon absorption profile, which can be written as [39]

$$G(z) = \int_{\omega_g}^{\infty} G^{abs}(\omega)/z_c \, d\omega \quad (2)$$

, where $\omega_g$ is the angular frequency corresponding to the band gap of the thermionic emitter, $G^{abs}$ is the monochromatic photon flux generated by the thermal radiator, which is absorbed in the thermionic emitter and $z_c$ is the thickness of the thermionic emitter. This photon flux is calculated with the help of fluctuational electrodynamics. Fluctuational electrodynamics is a first-principles approach to calculate the thermal radiation. It accounts for the near-field coupling of evanescent photons as well as the far-field propagating photons (also described by the Stefan-Boltzmann law). According to the fluctuation-dissipation theorem [40], thermal radiation in a medium is caused by the random charge fluctuations inside the medium. These fluctuations result in a space-time-dependent electric current density (with a zero time average), which then creates fluctuating electromagnetic fields. Fluctuational electrodynamics models this effect by adding a stochastic current density to Maxwell's equations. The fluctuation-dissipation theorem provides the relationship between the stochastic current source and the local temperature of the radiating medium [40,41]. The radiative heat or photon flux at a given wavelength at a particular position within the absorber is determined by calculating the component of the time-averaged Poynting vector perpendicular to the irradiated medium's surface. $G_\omega^{abs}$ is obtained by calculating the difference between the photon fluxes, $\varphi_{photon}(\omega,z,T)$, reaching the thermionic emitter from the thermal radiator and leaving the thermionic emitter toward the electron collector [42]:

$$G_\omega^{abs} = \varphi_{photon}(\omega, z = G_p^+, T_s) - \varphi_{photon}(\omega, z = G_p + z_c^+, T_s) \quad (3.a)$$

, where

$$\varphi_{photon}(\omega,z,T_s) = \frac{k_v^2 \Theta(\omega,T_s)}{2\pi^2} \text{Re}\left\{\frac{i\varepsilon_{rs}^"(\omega)}{\hbar\omega}\int_0^\infty \frac{\beta d\beta}{k_s}\left[\begin{array}{l}g_{sl\rho\rho}^E(\beta,z,\omega)g_{sl\theta\rho}^{H*}(\beta,z,\omega)+\\ g_{sl\rho z}^E(\beta,z,\omega)g_{sl\theta z}^{H*}(\beta,z,\omega) - g_{sl\theta\theta}^E(\beta,z,\omega)g_{sl\rho\theta}^{H*}(\beta,z,\omega)\end{array}\right]\right\}.$$

(3.b)

In the above equations, $\omega$ is the angular frequency, $k_v$ is the vacuum wavevector, $\Theta(\omega,T)$ is the mean energy of a Planck oscillator at frequency $\omega$ and temperature $T$, $\varepsilon_{rs}^"$ is the imaginary component of the dielectric constant in layer s, $\beta$ and $k_s$ are the parallel and perpendicular components, respectively, of the wavevector in medium s, and $\hbar$ is the reduced Planck's constant. The subscripts s and l correspond to the source (thermal radiator) and absorber medium (thermionic emitter), respectively. $\rho,\theta$ and $z$ are the coordinates as shown in fig. 1 and $G_p$, $z_c$ are the photon coupling gap width and the thickness of the thermionic emitter, respectively. The terms $g_{sl}^E$ and $g_{sl}^H$ represent, respectively, the Weyl components of the electric and magnetic dyadic Green's functions (DGFs). These are spatial transfer functions relating the fields in medium l at position with frequency $\omega$ and wavevector $\beta$ to a source located in medium s. The explicit mathematical formulation of the above-mentioned Weyl components of the DGFs, as well as the numerical procedure to solve eq. (3), can be found in [42]. Since eq. (3.b), normalized by the mean energy of the Planck oscillator, solely depends on the geometry and material properties of the radiatively coupled system [42,43], the monochromatic photon flux which is thermally radiated by the thermionic emitter and absorbed by the thermal radiator can be calculated by replacing the thermal emitter temperature $T_s$ with the thermionic emitter temperature $T_E$ in eq. (3).

In solving the continuity eq. (1), one needs boundary conditions at the two surfaces of the thermionic emitter. At both surfaces, the electrons can recombine with the surface defects and this effect is modelled by a surface recombination velocity. Also, at the surface facing the collector, the electrons can leave through thermionic emission. Therefore, the following boundary conditions are used to model these surface phenomena [44,45]:

$$D_n \frac{dn}{dz}\Big|_{z=0} = S_{n,F}[\delta n(0)] \quad (4.a) \quad \text{and} \quad D_n \frac{dn}{dz}\Big|_{z=z_c} = -S_{n,B}[\delta n(z_c)] - J_{net}/e \quad (4.b)$$

, where $S_{n,F}$ and $S_{n,B}$ are the surface recombination velocities at the front and back surfaces of the thermionic emitter, respectively, $\delta n(z) = n(z) - n_{eq}(z)$ is the excess electron concentration and $J_{net}$ is the emitted net current density. The non-equilibrium radiative recombination rate is calculated as [46]

$$R_{Radiative} = \int_{\omega_g}^{\infty} (\frac{np}{n_{eq} p_{eq}} - 1) \frac{\varphi_{photon}(\omega, z=0^-, T_E)}{z_c} d\omega \quad (5)$$

, where $n$ and $p$ are the steady-state electron and hole concentrations, respectively and $n_{eq}$ and $p_{eq}$ correspond to their equilibrium values.

The Auger recombination rate is given by [47]

$$R_{Auger} = C_n(n^2 p - n_{eq}^2 p_{eq}) + C_p(np^2 - n_{eq} p_{eq}^2) \quad (6)$$

, where $C_n$ and $C_p$ are the electron and hole Auger recombination coefficients, respectively. The Shockley-Reed-Hall (SRH) recombination rate follows [47]

$$R_{SRH} = \frac{np - n_i^2}{\tau_p(n+n_1) + \tau_n(p+p_1)} \quad (7)$$

, where $n_1 = n_{eq} e^{\frac{E_T - E_F}{k_B T_E}}$, $p_1 = p_{eq} e^{\frac{E_F - E_T}{k_B T_E}}$ and $n_i$ is the intrinsic carrier concentration. In the above equations, $E_F$ is the Fermi level and $E_T$ is the trap energy level, which we consider to be at the intrinsic energy level (*i.e.*, at the mid gap). $\tau_n$ and $\tau_P$ are the SRH recombination lifetimes for electrons and holes, respectively. The above-mentioned detailed recombination models are valid under both low- and high-injection levels.

### 2.1.3. Space charge effect in the electron coupling gap

The space charge effect in the electron coupling gap results from the coulombic repulsion of the thermionically emitted electrons. This charge repulsion results in an additional potential barrier in the vacuum gap, which the electrons need to overcome in order to reach the collector. This potential barrier can be obtained from the phase space analysis of the thermionically emitted electrons. This analysis involves solving the coupled Poisson-Vlasov equations [48–51]. The Vlasov equation provides the velocity distribution of the electrons in the electron coupling gap, which can be written as [48]

$$f(x, v_e) = 2n(x_m)\sqrt{m_e^3/(8\pi^3 k_B^3 T_E^3)} \exp\left(\frac{\varphi_m - \varphi(x) - \frac{1}{2}m_e v_e^2}{k_B T_E}\right) \Theta(v_{ex} \mp v_{ex,min}) \quad (8)$$

, where $\varphi_m$ and $\varphi_x$ are the maximum (see fig. 1b) and local motives in the interelectrode space, respectively, $m_e$ is the electron mass, $k_B$ is the Boltzmann constant, $T_E$ is the emitter temperature, $v_e$ and $v_{ex}$ are the electron velocity and its component perpendicular to the emitting surface, respectively, $n(x_m)$ is the electron density at the position of maximum motive, and $\Theta$ is the Heaviside step function. The upper and lower signs apply for $x > x_m$ and $x \leq x_m$, respectively. By integrating the electron distribution over the entire velocity space, we obtain the spatial distribution of charge carriers inside the vacuum gap as [48]

$$n(x) = \int_{-\infty}^{+\infty} dv_z \int_{-\infty}^{+\infty} dv_y \int_{-\infty}^{+\infty} dv_x f(x, v_e) = n(x_m) \exp(\gamma)[1 \mp \text{erf}(\sqrt{\gamma})] \quad (9)$$

, where $\gamma = (\varphi_m - \varphi(x))/k_B T_E$ is the dimensionless potential barrier and $\text{erf}(z) = \frac{2}{\sqrt{\pi}} \int_0^z \exp(-t^2) dt$ is the error function. The upper and lower signs apply for $x > x_m$ and $x \leq x_m$, respectively. The resulting space charge barrier is calculated by solving the Poisson equation, which can be written as [48]

$$2\frac{d^2\gamma}{d\xi^2} = \exp(\gamma)[1 \pm \text{erf}(\sqrt{\gamma})] \quad (10)$$

, where $\xi$ is the dimensionless position variable given by $\xi = (x - x_m)/x_0$, in which $x_0$ is the normalization length. The upper and lower signs apply for $x < x_m$ and $x \geq x_m$, respectively. The resulting emitter and collector current considering the space charge effect can be defined, respectively, as [50]

$$J_E = A_R T_E^2 e^{[-\frac{\varphi_m - (E_{F,n} - E_{F,E})}{k_B T_E}]} \quad (11.a) \quad \text{and} \quad J_C = A_R T_C^2 e^{(-\frac{\varphi_m - eV}{k_B T_C})}. \quad (11.b)$$

In the above equations, $A_R$ is the Richardson constant, $k_B$ is the Boltzmann constant and $E_{F,n} - E_{F,E}$ is the reduction in the electron emission barrier caused by the photogeneration-induced upshift of the electron quasi-Fermi level. Further details regarding the implementation of the above space charge model can be found in [51]. The net energy flux carried by the thermionic current from the emitter is given by [52]

$$Q_T = \frac{[(J_E - J_C)\varphi_m + 2k_B(T_E J_E - T_C J_C)]}{e}. \quad (12)$$

### 2.1.4. Semiconductor material model

Here, we discuss the relevant theories to determine various material parameters and carrier concentrations needed for the TPT device analysis. The equilibrium Fermi level in the emitter can be calculated from the charge neutrality criterion, which for a p-type semiconductor can be written as [47]

$$N_{\mathrm{C}} \, e^{-\frac{E_{\mathrm{g}}-E_{\mathrm{F}}}{k_{\mathrm{B}}T_{\mathrm{E}}}} + N_{\mathrm{A}} \frac{1}{1+4 \, e^{\frac{E_{\mathrm{A}}-E_{\mathrm{F}}}{k_{\mathrm{B}}T_{\mathrm{E}}}}} = N_{\mathrm{V}} \, e^{-\frac{E_{\mathrm{F}}}{k_{\mathrm{B}}T_{\mathrm{E}}}} \quad (13)$$

, where $N_{\mathrm{C}}$ and $N_{\mathrm{V}}$ are the effective densities of states in the conduction and valance band, respectively, $E_{\mathrm{g}}$ is the band gap of the semiconductor, $N_{\mathrm{A}}$ is the doping concentration and $E_{\mathrm{A}}$ is the dopant energy level. The temperature dependencies of the effective densities of states and bandgap are given by [47]

$$N_{\mathrm{C}} = 2(\frac{2\pi m_n^* k_{\mathrm{B}}T}{h^2})^{3/2}, \quad (14.\mathrm{a})$$

$$N_{\mathrm{V}} = 2(\frac{2\pi m_p^* k_{\mathrm{B}}T}{h^2})^{3/2} \quad (14.\mathrm{b})$$

$$\text{and } E_{\mathrm{g}}(T) = E_{\mathrm{g}0} - aT^2/(T+b) \quad (14.\mathrm{c})$$

, where $m_n^*$ and $m_p^*$ are the density of states effective masses of the semiconductor, $E_{\mathrm{g}0}$ is the band gap at 0 K, $h$ is the Planck constant and $a,b$ are the empirical fitting parameters of the band gap narrowing effect. Once the equilibrium Fermi level is determined, the equilibrium electron and hole concentrations can be calculated as [47]

$$n_{\mathrm{eq}} = N_{\mathrm{C}} \, e^{-\frac{E_{\mathrm{g}}-E_{\mathrm{F}}}{k_{\mathrm{B}}T_{\mathrm{E}}}} \quad (15.\mathrm{a}) \text{ and } p_{\mathrm{eq}} = N_{\mathrm{V}} \, e^{-\frac{E_{\mathrm{F}}}{k_{\mathrm{B}}T_{\mathrm{E}}}}. \quad (15.\mathrm{b})$$

Similarly, the emitter work function is related to the Fermi level as

$$\varphi_{\mathrm{E}} = E_{\mathrm{g}} - E_{\mathrm{F}} + \chi \quad (16)$$

, where $\chi$ is the electron affinity of the emitter. The electron diffusion coefficient, $D_{\mathrm{n}}$, is calculated using the general relationship [47]

$$D_{\mathrm{n}} = \frac{k_{\mathrm{B}}T_{\mathrm{E}}\mu_{\mathrm{n}}}{e} F_{1/2}(\frac{E_{\mathrm{F}}-E_{\mathrm{C}}}{k_{\mathrm{B}}T_{\mathrm{E}}}) / F_{-1/2}(\frac{E_{\mathrm{F}}-E_{\mathrm{C}}}{k_{\mathrm{B}}T_{\mathrm{E}}}) \quad (17)$$

, where $\mu_n$ is the electron mobility and $F_n$ is the Fermi-Dirac integral of order n. Under low injection levels, the above equation reduces to Einstein's relation. The amount of optical upshift in the electron quasi-Fermi level can be written as [47]

$$E_{F,n} - E_{F,E} = k_B T_E \ln(n/n_{eq}). \quad (18)$$

**2.1.5. Self-consistent solution of the electrode thermal balance**

As the TPT performance strongly depends on the electrode temperatures, it is crucial to calculate these temperatures and their dependencies on the incident radiation flux, the device material properties, and the operating conditions. The emitter thermal balance is calculated considering various mechanisms by which heat is removed from the emitter by electrons and photons. This can be written as

$$Q_{In} = Q_{Noneq} + Q_{Eq} + Q_{Rad} + Q_T + Q_{Lead(Thermal\ Conduction+Joule\ Heating)}. \quad (19)$$

In this equation, $Q_{In}$ is the energy absorbed by the thermionic emitter from the thermal radiator via photon coupling; $Q_{Noneq}$ is the radiative loss due to non-equilibrium radiative recombination; $Q_{Eq}$ is the loss due to equilibrium radiative recombination and sub-band gap thermal radiation processes; $Q_{Rad}$ is the net energy exchange between the thermionic emitter and collector via radiative coupling (which we also calculate using fluctuational electrodynamics and the procedure is similar to that described in section 2.1.2); $Q_T$ is the energy taken away from the thermionic emitter by electron emission and $Q_{Lead(Thermal\ Conduction+Joule\ Heating)}$ is the heat loss in the lead due to thermal conduction and Joule heating. Both $Q_{Eq}$ and $Q_{Noneq}$ are recycled in the thermal radiator.

The value of the lead resistance must be chosen carefully to minimize the lead-related loss as there exists a trade-off between the thermal conduction and Joule heating [49]: a low resistance would reduce the lead voltage drop but increase the thermal conduction loss (according to the Wiedemann–Franz law), and a high resistance would entail low thermal conduction loss but result in significant Joule heating and voltage drop in the leads. Considering these conflicting requirements, we used the optimal value of lead resistance for maximum conversion efficiency [49]:

$$\eta_{\text{Lead optimized}} = \frac{SJ_{\text{net}}(V - SJ_{\text{net}}R_{\text{Lead}})_{\text{Max}}}{SQ_{\text{Net}}} \quad (20)$$

, where $S$ is the cross-sectional area of the electrodes, $R_{\text{Lead}}$ is the lead resistance, $J_{\text{net}}$ is the net current density and $Q_{\text{Net}}$ $(= Q_{\text{In}} - Q_{\text{Noneq}} - Q_{\text{Eq}})$ is the net energy exchange between the thermal radiator and the thermionic emitter.

The heat flux which is absorbed by the collector and released to a heat sink held at room temperature is given by [53]

$$Q_{\text{Sink}} = K_L(T_C - T_0) = Q_{\text{Rad}} + Q_T + Q_{\text{Lead(Thermal Conduction+Joule Heating)}} - P_{\text{Elec}} \quad (21)$$

, where $K_L$ is the heat transfer coefficient between the collector and heat sink, $T_C$ and $T_0$ are the collector and heat sink temperatures, respectively and $P_{\text{Elec}}$ is the useful electrical output of the TPT device. These thermal balance conditions, in addition to the coupling between the emitter particle balance and the space charge effect, are implemented in our model using a self-consistent iterative algorithm (see fig. S1 in the supplementary document).

## 2.2. Thermophotovoltaic device modeling

Similarly, to the TPT device, the photogeneration of electron-hole pairs and the related transport of the minority carriers across the thermophotovoltaic device can be calculated using eq. (1). However, as the TPV device consists of regions with different types of doping (namely, p- and n-type) and a depletion layer at the junction of these dissimilarly doped regions, eq. (1) must be solved in both n- and p-type regions with appropriate boundary conditions. The photon absorption and the associated carrier generation profile in each of the abovementioned regions are calculated with the help of fluctuational electrodynamics and the scattering matrix method for layered media [42,43]. The procedure is similar to that described in section 2.1.2 and will not be repeated here. The boundary conditions at the edge of the p-region facing the thermal radiator and at the edge of the n-region connected to the TPV back-contact, respectively, are [43]

$$D_\text{n} \frac{dn}{dz}\Big|_{z=0} = S_{n,F}[\delta n(0)] \quad (22.\text{a}) \quad \text{and} \quad D_\text{p} \frac{dp}{dz}\Big|_{z=z_\text{TPV}} = -S_{n,B}[\delta p(z_\text{TPV})]. \quad (22.\text{b})$$

In the above equations, $z_\text{TPV}$ is the thickness of the photovoltaic material defined as $z_\text{TPV} = z_\text{p} + z_\text{dp} + z_\text{n}$, where $z_\text{p}$, $z_\text{dp}$ and $z_\text{n}$ are the thicknesses of the neutral p, depletion and neutral n regions, respectively. The diffusion coefficient for holes, $D_\text{p}$, can be calculated similarly as described above for the electrons.

At the edges of the depletion region, the minority carriers are swept by the electric field at the p-n junction, so the boundary conditions at these edges are [43]

$$\delta n(z_\text{p}) = \delta p(z_\text{p} + z_\text{dp}) = 0. \quad (23)$$

Once the spatial distribution of the charge carriers in the TPV device is known, the TPV photocurrent density, $J_\text{ph}$, is calculated. The contributions to this photocurrent from the p, n and depletion regions are, respectively, [39,43]

$$J_p = eD_\text{n} \frac{dn}{dz}\Big|_{z=z_\text{p}}, \quad (24.\text{a})$$

$$J_n = -eD_\text{p} \frac{dp}{dz}\Big|_{z=z_\text{p}+z_\text{dp}} \quad (24.\text{b})$$

$$\text{and} \quad J_\text{dp} = e\int_{Z_\text{p}}^{Z_\text{p}+Z_\text{dp}} G_\text{dp}(z)\, dz \quad (24.\text{c})$$

, where $G_\text{dp}(z)$ is the local photogeneration rate of electron-hole pairs in the depletion region. In the above calculations, the width of the depletion region is determined as [43]

$$z_\text{dp} = \left[\frac{2\varepsilon_s V_\text{bi}}{e}\left(\frac{1}{N_A} + \frac{1}{N_D}\right)\right]^{1/2} \quad (25)$$

, where $V_\text{bi}$ is the built-in voltage across the depletion region, $\varepsilon_s$ is the static permittivity of the photovoltaic material and $N_A$ and $N_D$ are the doping concentrations of the p and n regions, respectively.

The photocurrent (defined as the total current of the TPV device due to illumination) is given by [39]

$$J_{ph} = J_p + J_n + J_{dp}. \quad (26)$$

The net current of the TPV device can be calculated by subtracting the dark current from the photocurrent *i.e.,* $J_{net}(V_o) = J_{ph} - J_{dark}(V_o)$, where $V_o$ is the operating voltage. The dark current is found in a similar manner as described above for the photocurrent except that the photogeneration term is set to zero and the following boundary conditions at the edges of the depletion region are used [43]:

$$\delta n(z_p) = n_{eq} e^{\frac{eV_o}{k_B T_{TPV}}} \text{ (27.a) and } \delta p(z_p + z_{dp}) = p_{eq} e^{\frac{eV_o}{k_B T_{TPV}}} \text{ (27.b)}.$$

A practice found in the TPV literature is to use a wide-base diode approximation in the calculation of the dark current [39,54,55]; however, our approach is free of such an assumption, making it more general and physically accurate.

## 2.3. Model Validation

The various models that are described in the above sections have been validated by benchmarking them against the experimental and theoretical findings in the existing literature. Some of the benchmarking results are shown in the supplementary document (see fig. S2-S4).

## 3. Results and Discussion

First, we discuss our choice of materials for this study. For the thermal radiator, we chose hexagonal boron nitride (hBN) due to its intrinsic spectral selectivity–it supports surface phonon polaritons which result in a sharp peak in the radiation spectrum at their resonance frequency [42,56]. However, due to the evanescent nature of these surface modes, they can be coupled only if the thermal radiator is in close vicinity of the thermionic emitter (*i.e.,* for nanoscale photon coupling gaps). The strength and nature of this polaritonic enhancement of radiative coupling strongly depend on both of the coupled materials' ability to support surface modes as well as the resonance frequency of these supported modes [57] (see fig. 2(a)). Therefore, to maximize thermo-photon coupling, the photon-to-electron converter (*i.e.* the thermionic electron emitter) would ideally use a material with dielectric properties similar to that of the thermal radiator. For example, in some TPV studies, photovoltaic materials or thermal radiators with fictitious dielectric properties are assumed to maximize the thermo-photon coupling [54,58]. However, such an assumption, while interesting for predicting the upper limit of the device performance, is unrealistic. Therefore, for a realistic performance estimate, we choose indium arsenide (InAs) as the photon-to-electron conversion material. InAs has a high melting point, which makes it suitable for thermionic conversion applications [59]. Also, the band gap of InAs is suitable to couple the surface modes generated by hBN under the TPT device's operating conditions (see fig. 2(b)). Moreover, InAs is widely considered for thermophotovoltaics applications [23,60–62], which is another motivation to use this material for comparison between thermo-photo-thermionics and thermophotovoltaics.

The optical properties of the thermal radiator and thermionic emitter are taken from [56,63]. The parameters related to various recombination mechanisms, the electron and hole mobilities, the empirical fitting parameters for the band gap narrowing effect and the electron and hole effective masses of InAs are taken from [59,64–66]. For the TPT device's thermionic emitter, we assume p-type doping with a concentration of $10^{18}$ cm$^{-3}$ and the dopant energy level of zinc impurity [59], a thickness of 2 µm (A higher thickness was found not to make any significant changes in the radiative coupling throughout the device.), and an electron affinity of 1 eV, which may be achieved by coating the emitter surface with a low work function material such as cesium, doped diamond, etc. For the metallic collector, we consider a work function of 1 eV (also achievable via low work function material coating) and the dielectric properties of tungsten [67]. For heat removal from the collector, we assume a heat transfer coefficient of 0.1 Wcm$^{-2}$ K$^{-1}$ between the collector and heat sink. For the TPV device, we consider a p-on-n configuration, which is preferred over n-on-p to prevent short-circuiting at the junction caused by the annealing of the ohmic contact to the doped region [39,43]. The doping concentrations and thicknesses of the p and n regions in the TPV device are taken from the literature [39,43], with the n-type dopant energy level of germanium impurity [59]. For both the TPT and TPV devices, we investigate the device operation for a wide range of the photon coupling gap size spanning the near-field to the far-field limit. For ease of access, the important material and device parameters are given in supplementary table 1.

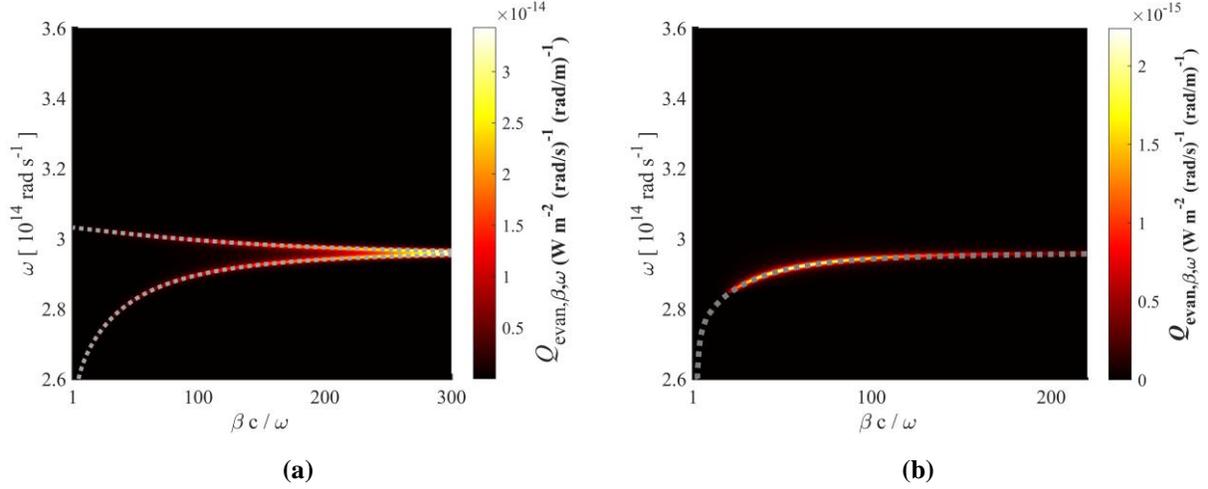

**Fig. 2.** Evanescent radiative flux per unit angular frequency, $\omega$, and per unit parallel wavevector, $\beta$, between **(a)** two hBN half-spaces and **(b)** a hBN half-space and an InAs thin film as a function of the angular frequency and the dimensionless parallel wavevector component. **c** is the speed of light. Superimposed on the radiative fluxes are the surface-polariton dispersion relations (shown as grey dotted lines) in TM polarization. In fig. (a), the lower- and higher-frequency branches of the dispersion relation correspond to the symmetric and anti-symmetric modes, respectively [68]. The data are shown for a vacuum gap size of 10 nm and a temperature difference of 200 K, with the temperature of the hBN half-space thermal radiator being 1100 K.

To discuss the merits of the TPT device, in fig. 3, we show the power density and conversion efficiency at both the maximum-power-point (MPP) and the maximum-efficiency-point (MEP). These two points in the TPT device are different because the radiative exchange between the thermal radiator and the thermionic emitter, as well as the various energy exchange channels within the thermionic device (which determine the electrode temperatures), have complex dependencies on the electron and photon coupling gap sizes and the operating voltage of the device [53]. Also in fig. 3, we show the current density and the corresponding optimal output voltage at both MPP and MEP for their respective optimal electron coupling gap sizes. We observe that the power and current densities decrease with an increase in the photon coupling gap size, due to the weakening of the radiative coupling between the thermal radiator and thermionic emitter.

However, the conversion efficiency and optimal output voltage both increase with the photon coupling distance. To explain the non-monotonic trend in the MEP efficiency curve (around photon coupling gaps of 50 nm and 500 nm) below, in table 1, we show the input heat flux and various energy exchange channels within the TPT converter for the relevant photon coupling gap sizes. In table 2, we show the changes in various energy exchange channels at MEP as the photon coupling gap size changes around those values.

**Table 1: Various energy exchange channels within the TPT converter for the different photon coupling gap sizes around the two peaks of the MEP efficiency curve on fig. 3(a).**

| Photon coupling gap size (nm) | Net radiative exchange across the photon coupling gap (Input heat flux) (W cm$^{-2}$) | Net radiative exchange across the electron coupling gap (W cm$^{-2}$) | Net thermionic exchange across the electron coupling gap (W cm$^{-2}$) | Loss across the electrical lead (W cm$^{-2}$) |
|---|---|---|---|---|
| 30 | **3.467** | 0.124 | **2.864** | 0.403 |
| 50 | **3.016** | 0.122 | **2.738** | 0.272 |
| 70 | **2.384** | 0.116 | **1.960** | 0.242 |
| 300 | **1.528** | 0.106 | **1.267** | 0.134 |
| 500 | **1.013** | 0.108 | **0.861** | 0.077 |
| 700 | **0.914** | 0.103 | **0.747** | 0.058 |

**Table 2: Decrease in different energy exchange channels within the TPT converter for transitions between photon coupling gap sizes shown on table 1.**

| Transition in photon coupling gap size (nm) | Decrease in radiative exchange across the photon coupling gap (Input heat flux) (W cm$^{-2}$) | Decrease in radiative exchange across the electron coupling gap (W cm$^{-2}$) | Decrease in thermionic exchange across the electron coupling gap (W cm$^{-2}$) | Decrease in loss across the electrical lead (W cm$^{-2}$) |
|---|---|---|---|---|
| 30 to 50 | **0.451** | 0.002 | **0.126** | 0.131 |
| 50 to 70 | **0.632** | 0.006 | **0.778** | 0.030 |
| 300 to 500 | **0.515** | -0.002 | **0.406** | 0.057 |
| 500 to 700 | **0.099** | 0.005 | **0.114** | 0.019 |

As can be seen in table 2, going from 30 nm to 50 nm, the input heat flux decreases by 0.451 W cm$^{-2}$, thermionic exchange decreases by 0.126 W cm$^{-2}$ and the net decrease in other loss channels is 0.133 W cm$^{-2}$. Therefore, during this transition, the input heat flux decreases at a much faster rate than that of the net decrease in thermionic exchange, which naturally leads to an increase in conversion efficiency. On the other hand, during the transition from 50 to 70 nm, the input heat flux decreases by 0.632 W cm$^{-2}$ and the thermionic exchange decreases by 0.778 W cm$^{-2}$. Therefore, during this transition, the rate of decrease of the input heat flux is lower than the rate of the decrease in thermionic exchange and other loss channels, leading to a decrease in the conversion efficiency. The other peak can be explained by similar reasoning.

Shown in fig. 3(c) are the trends in the electrode temperatures at both MEP and MPP as a function of the photon coupling gap size. All curves exhibit a downward trend with the increase of photon coupling gap size except for the MEP emitter temperature, which exhibits a local maximum due to the complex interplay among various energy exchange channels (*e.g.*, as seen in the tables above) between the electrodes [53,67].

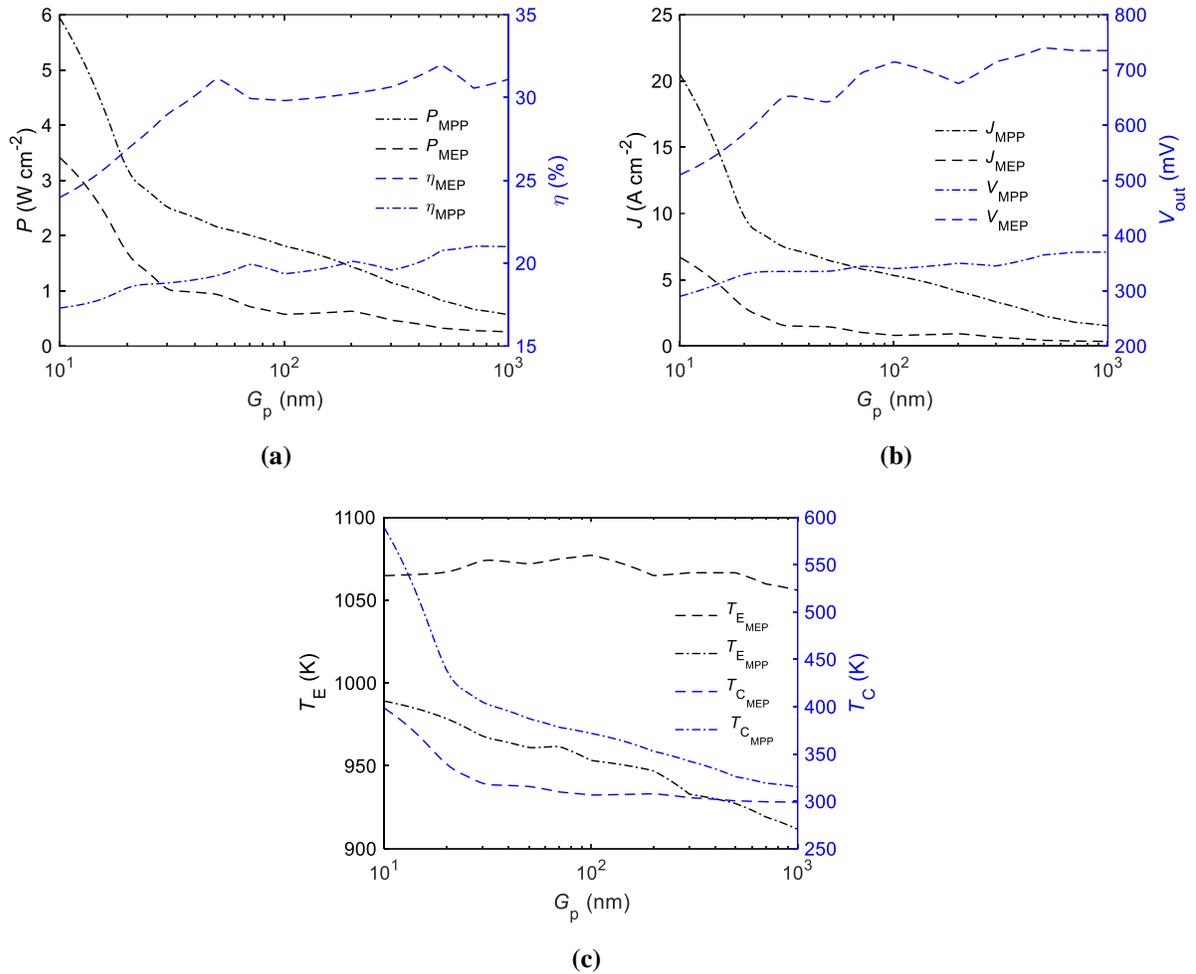

**Fig. 3 (a)** Output power density and conversion efficiency, **(b)** Current density and output voltage and **(c)** Electrode temperatures of the TPT device as a function of the photon coupling gap size. The dash-dotted lines represent the data at MPP and the dashed lines represent the data at MEP. The curves are shown for the optimal electron coupling gap width and a thermal radiator temperature of 1100 K.

We thus observe that maximizing both the output power density and conversion efficiency of the TPT device are somewhat contradictory in terms of the photon coupling gap size (which partly determines the electrode temperatures) and a more efficient operation often involves lower electrode temperatures. A crucial and counterintuitive observation is that the highest efficiency and maximum electron emitter temperatures are not achieved simultaneously by simply bringing the electron emitter to physical contact with the thermal radiator, and that the inclusion of a photon tunneling gap enables a useful degree of freedom for performance optimization.

To further elucidate the complex dependencies of the device performance on photon and electron coupling strengths, the variations of the power and current densities, electrode temperatures and conversion efficiency with the photon and electron coupling gap sizes are shown in figs. 4 and 5 for MEP and MPP, respectively. In all cases, the value of output voltage that maximizes the TPT performance at each combination of gap sizes is used.

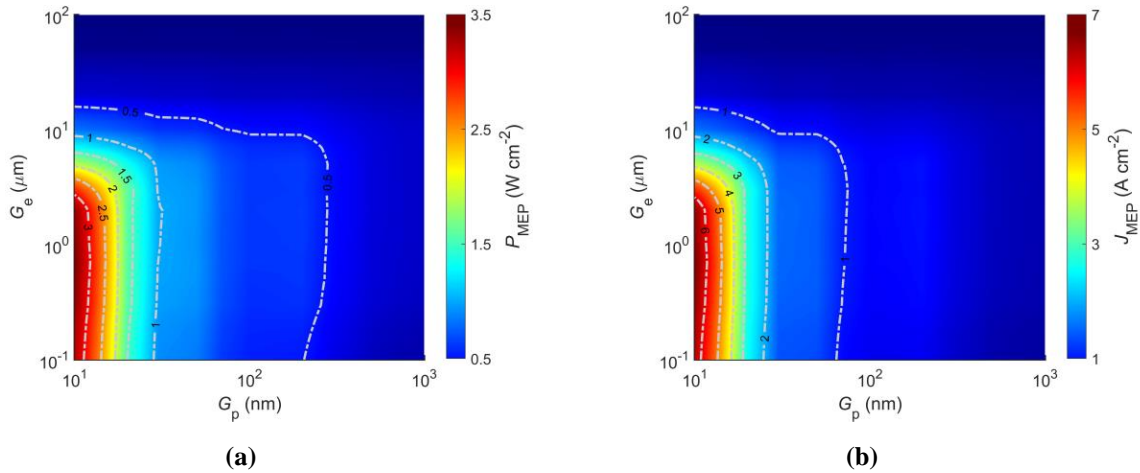

(a)  (b)

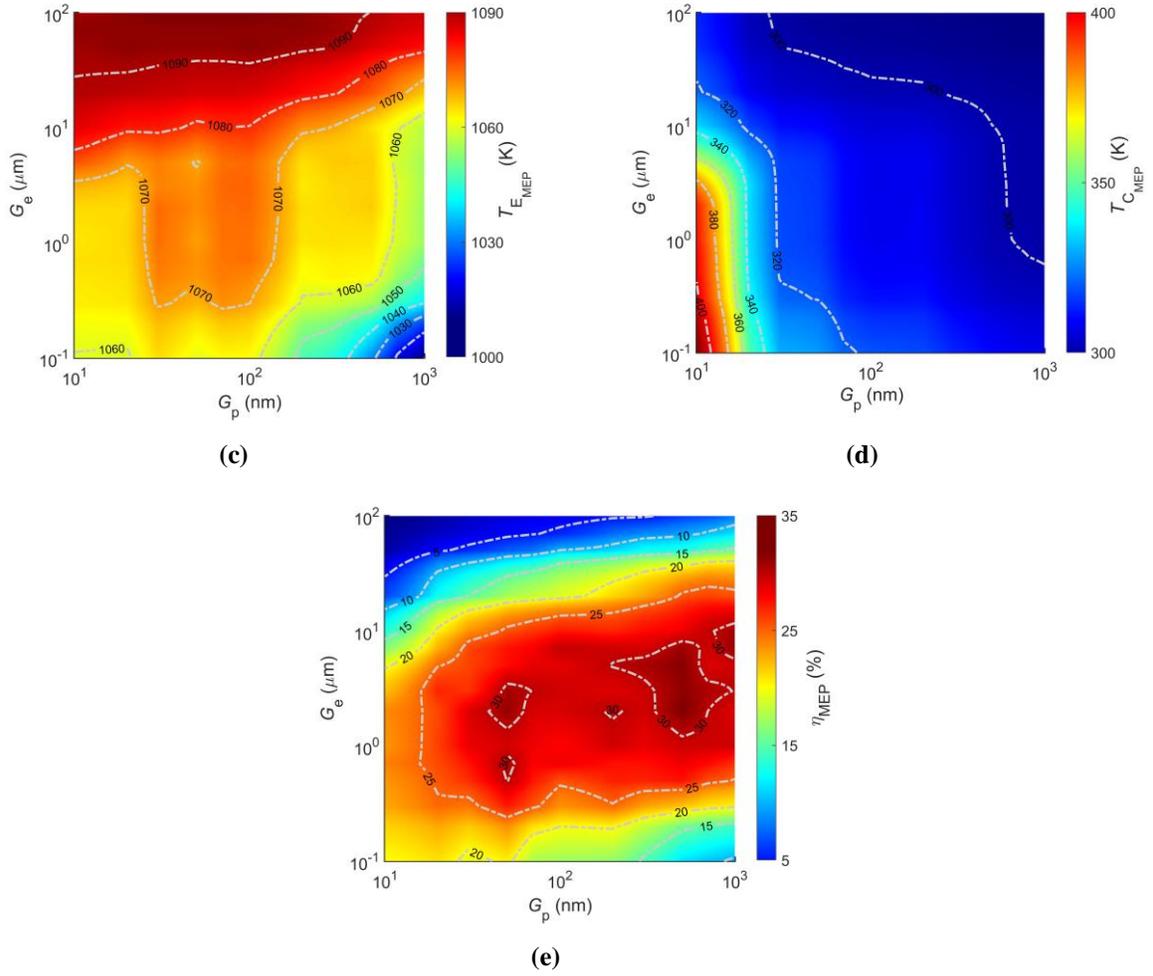

**Fig. 4.** **(a)** Output power density, **(b)** Current density, **(c)** Emitter temperature, **(d)** Collector temperature and **(e)** Conversion efficiency as a function of the photon coupling (x-axis) and electron coupling (y-axis) gap sizes. The graphs are for MEP, an output voltage that optimizes the conversion efficiency and a thermal radiator temperature of 1100 K.

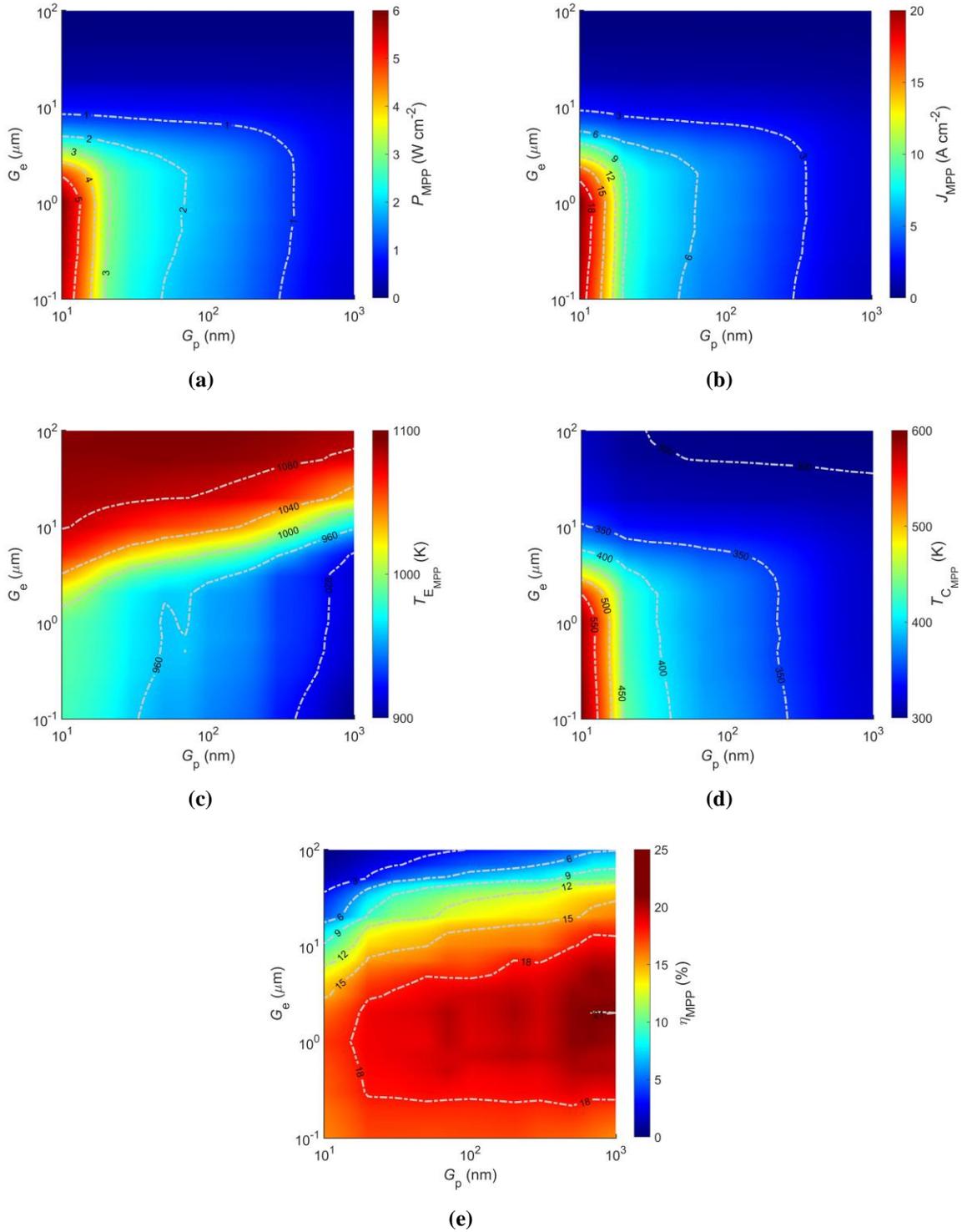

**Fig. 5. (a)** Output power density, **(b)** Current density, **(c)** Emitter temperature, **(d)** Collector temperature and **(e)** Conversion efficiency as a function of the photon coupling (x-axis) and electron coupling (y-axis) gap sizes. The graphs are for MPP, an output voltage that optimizes the output power density and a thermal radiator temperature of 1100 K.

We note that for both MPP and MEP, all the displayed quantities have similar trends with the electron coupling gap size. For example, for a given photon coupling gap size, the variations of the power density, current density and the conversion efficiency all exhibit local maxima with the electron coupling gap size. These trends result from a trade-off between the space-charge loss and the photon tunneling enhancement of radiative coupling between the electron emitter and collector. Similarly, the emitter and collector temperatures increase and decrease, respectively with the increase of the electron coupling gap size. This is due to the reduction in the thermionic current (due to increased space charge) and interelectrode radiative coupling strength as the electron coupling gap size increases [67].

Now, we investigate the performance of a thermophotovoltaic converter with thermal radiator and semiconductor materials identical to those used for the TPT converter. For this analysis, we use favorable conditions for the TPV device by assuming that it has no series resistance and that it can be maintained around room temperature irrespectively of its operating conditions. This will enable conservative estimates for the comparative advantages of the TPT device.

In figs. 6 (a) and (b), we show the output power density, conversion efficiency, current density and optimal output voltage for the TPV device for a wide range of thermal radiator temperatures. (Note that, for the TPV device, MPP and MEP are the same under the favorable conditions mentioned above.) The conversion efficiencies shown here are consistent with the TPV literature for similar material band gaps [23]. The plotted quantities' downward trend with the increasing photon coupling gap size is expected based on the dependence of the photon-coupling strength on the distance between the thermal radiator and the PV material. It is also worth discussing the cooling requirements of the TPV device (fig. 6(c)) to obtain the plotted performance metrics. To maintain the TPV device at 300 K (with the heat sink at 293 K), the heat transfer coefficient between the device and the heat sink needs to exceed the upper limit of free convection [69] (the value we used for cooling the collector of the TPT device) even in the far-field limit for the thermal radiator temperatures considered. Therefore, the cooling of the TPV device requires additional power and the actual conversion efficiency for this device is lower than the estimates shown here.

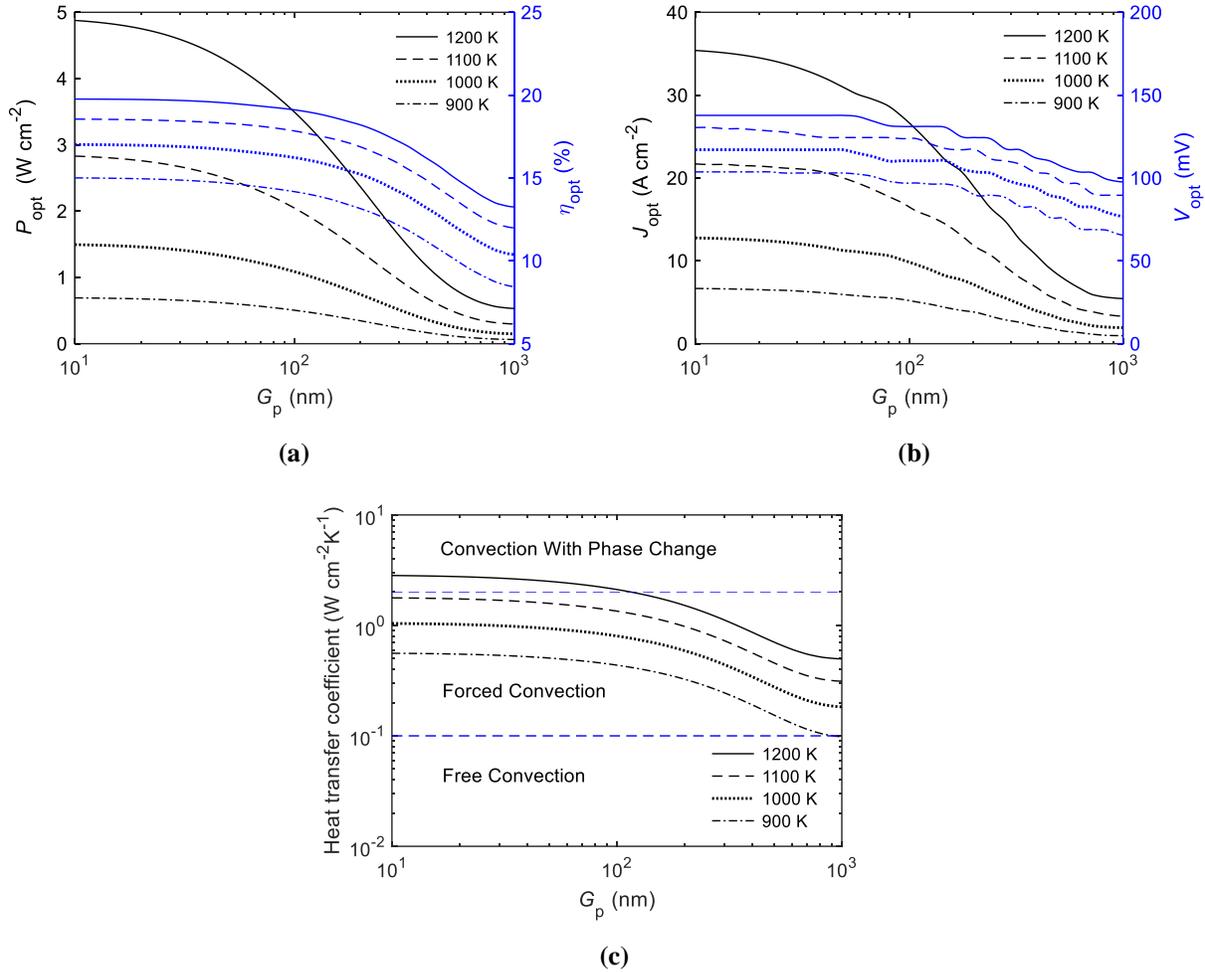

**Fig. 6.** **(a)** Output power density and conversion efficiency, **(b)** Current density and output voltage and **(c)** Cooling load of the TPV device as a function of the photon coupling gap size. The data are shown at the optimal operating point and for different thermal radiator temperatures.

We now present an illustrative comparison of the performance estimates between the TPT and TPV mechanisms (table 1). Most notably, the optimal output voltage in the TPT device can exceed the band gap of InAs, which is due to the thermal boost given to the electrons in the thermionic emission process. As well, the TPT device is superior to the TPV device in terms of power density and conversion efficiency. All these results indicate that TPT could be a promising alternative to the TPV mechanism for thermal and optical energy harvesting.

Finally, we note that the results shown in this study for the thermionic and thermophotovoltaic devices by no means define upper limits of conversion performance achievable for either one. The performance of these devices depends on the design of the thermal radiator and the heterostructure configuration used for electrical power generation. The purpose of the comparisons presented here is to demonstrate the relative advantages of the thermionic mechanism over thermophotovoltaics for identical material properties.

**Table 3: Performance comparison between the TPT and TPV converters for a photon coupling gap size of 10 nm and a thermal radiator temperature of 1100 K**

| Performance Metrics | TPT | TPV |
|---|---|---|
| Power Density (W cm$^{-2}$) | MPP: 5.94 <br> MEP: 3.42 | 2.8 |
| Current Density (A cm$^{-2}$) | MPP: 20.5 <br> MEP: 6.70 | 21.7 |
| Conversion Efficiency (%) | MPP: 17.3 <br> MEP: 24 <br> Note: Considering cooling power | 18.5 <br> Note: Neglecting cooling power |
| Output Voltage (mV) | MPP: 290 <br> MEP: 510 <br> Note: Exceeds material band gap near room temperature | 130 <br> Note: Not considering series resistance |
| Photon tunneling enhancement of power density over two decades of photon coupling gap range | MPP: 10.6 <br> MEP: 13.2 | 9.4 |
| Required Cooling Load | Upper limit of free convection | Exceeds upper limit of free convection |
| Hybridization Prospect: Carnot Potential (%)** | MPP: 50.25 <br> MEP: 26.46 | 2.33 |

** Carnot Potential is defined as the efficiency of a Carnot engine.

## 4. Summary, Conclusion and Outlook

In summary, we presented an alternative approach to harvesting thermo-photons using thermionic emission and compared it to thermophotovoltaics. We observed that the thermo-photo-thermionic route can outperform thermophotovoltaics in several key conversion performance metrics such as power density and conversion efficiency. Most notably, the output voltage in the TPT device can exceed the semiconductor material's bandgap, which sets a fundamental upper limit on the output voltage in the TPV mechanism. Crucially, we showed that optical coupling to the thermal source can enable higher efficiency than physical contact, making TPT a useful mechanism from the application's point of view.

Our results indicate that photon-coupled thermionic energy harvesting is a viable path and could be potentially superior to thermophotovoltaics. A TPT device could serve a wide variety of applications ranging from renewable energy harvesting and recycling waste heat to powering autonomous robots for expeditions to other planets. For example, the proposed device can be utilized for solar thermal conversion; recently, the global installed capacity of such solar thermal plants has been reported to be around 6.4 GW [70]. As another example, the TPT device can be used to recycle the high temperature waste heat from iron and steel industries where a waste heat temperature of around $1000^0$ C is available [71].

Precise nanoscale gaps have already been demonstrated in thermophotovoltaic devices [23], and similar fabrication technologies can be used to implement the photon tunneling gap in TPT devices. The required electron coupling gap size is of the order of a few micrometers under optimal operation. This is achievable, for instance, by using mechanically robust and thermally insulating spacers made using the atomic layer deposition technique [72]. Thermally stable low work function coatings are needed to lower the vacuum barrier for thermionic emission. As an example, such a low work function coating can be obtained from a phosphorus-doped diamond film with a reported work function as low as 0.9 eV and thermal stability at temperatures above 1000 K [73]. Combining all these technologies represents significant engineering challenges, but does not involve fundamental limitations. Thus, the promise of such high power density with reasonable efficiency allows us to envision chip-scale power plants. Additionally, the high temperature nature of the TPT converter means that it can be hybridized with TPV devices [74,75], making the two devices complementary rather than being in competition.

**Supplementary Material**

See supplementary material for the values of different material and device related parameters, self-consistent iterative algorithm and validation of the computational models used in this work.

**Acknowledgements**

We acknowledge funding from the Natural Sciences and Engineering Research Council of Canada (Grants No. RGPIN-2017-04608 and No. RGPAS-2017-507958). This research was undertaken thanks in part to funding from the Canada First Research Excellence Fund, Quantum Materials and Future Technologies Program. Ehsanur Rahman thanks the Natural Sciences and Engineering Research Council of Canada for a Vanier Canada Graduate Scholarship and the University of British Columbia for an International Doctoral Fellowship and Faculty of Applied Science Graduate Award.

E.R. conceptualized the work; developed the algorithm; performed the simulation, model verification, and data visualization; and wrote the draft manuscript; A.N. provided technical guidance and reviewed and edited the manuscript.

**Data Availability**

The data that support the findings of this study are available from the corresponding author upon reasonable request.

**References**

[1] Forman C, Muritala IK, Pardemann R, Meyer B. Estimating the global waste heat potential. Renew Sustain Energy Rev 2016;57:1568–79. https://doi.org/10.1016/j.rser.2015.12.192.

[2] Coutts TJ. Review of progress in thermophotovoltaic generation of electricity. Renew Sustain Energy Rev 1999;3:77–184. https://doi.org/10.1016/S1364-0321(98)00021-5.

[3] Yugami H, Sasa H, Yamaguchi M. Thermophotovoltaic systems for civilian and industrial applications in Japan. Semicond Sci Technol 2003;18. https://doi.org/10.1088/0268-

1242/18/5/315.

[4]   Nelson RE. A brief history of thermophotovoltaic development. Semicond Sci Technol 2003;18. https://doi.org/10.1088/0268-1242/18/5/301.

[5]   Burger T, Sempere C, Roy-Layinde B, Lenert A. Present Efficiencies and Future Opportunities in Thermophotovoltaics. Joule 2020;4:1660–80. https://doi.org/10.1016/j.joule.2020.06.021.

[6]   Bitnar B, Durisch W, Holzner R. Thermophotovoltaics on the move to applications. Appl Energy 2013;105:430–8. https://doi.org/10.1016/j.apenergy.2012.12.067.

[7]   Teofilo VL, Choong P, Chang J, Tseng YL, Ermer S. Thermophotovoltaic energy conversion for space. J Phys Chem C 2008;112:7841–5. https://doi.org/10.1021/jp711315c.

[8]   Coutts TJ, Guazzoni G, Luther J. An overview of the fifth conference on thermophotovoltaic generation of electricity. Semicond Sci Technol 2003;18. https://doi.org/10.1088/0268-1242/18/5/302.

[9]   Anderson DJ, Wong WA, Tuttle KL. An overview and status of NASA's Radioisotope Power Conversion Technology NRA. Collect Tech Pap - 3rd Int Energy Convers Eng Conf 2005;3:1681–9. https://doi.org/10.2514/6.2005-5713.

[10]  Wernsman B, Mahorter RG, Siergiej R, Link SD, Wehrer RJ, Belanger SJ, et al. Advanced Thermophotovoltaic Devices for Space Nuclear Power Systems. AIP Conf. Proc., vol. 746, 2005, p. 1441.

[11]  Colangelo G, De Risi A, Laforgia D. Experimental study of a burner with high temperature heat recovery system for TPV applications. Energy Convers Manag 2006;47:1192–206. https://doi.org/10.1016/j.enconman.2005.07.001.

[12]  Woolf DN, Kadlec EA, Bethke D, Grine AD, Nogan JJ, Cederberg JG, et al. High-efficiency thermophotovoltaic energy conversion enabled by a metamaterial selective emitter. Optica 2018;5:213. https://doi.org/10.1364/optica.5.000213.

[13]  Bhatt R, Gupta M. Design and validation of a high-efficiency planar solar

thermophotovoltaic system using a spectrally selective emitter. Opt Express 2020;28:21869. https://doi.org/10.1364/oe.394321.

[14] Yang Y, Chang JY, Sabbaghi P, Wang L. Performance Analysis of a Near-Field Thermophotovoltaic Device with a Metallodielectric Selective Emitter and Electrical Contacts for the Photovoltaic Cell. J Heat Transfer 2017;139:1–9. https://doi.org/10.1115/1.4034839.

[15] Bhatt R, Kravchenko I, Gupta M. High-efficiency solar thermophotovoltaic system using a nanostructure-based selective emitter. Sol Energy 2020;197:538–45. https://doi.org/10.1016/j.solener.2020.01.029.

[16] Chen M, Chen X, Yan H, Zhou P. Theoretical design of nanoparticle-based spectrally emitter for thermophotovoltaic applications. Phys E Low-Dimensional Syst Nanostructures 2021;126:114471. https://doi.org/10.1016/j.physe.2020.114471.

[17] Chubb DL. Light Pipe Thermophotovoltaics (LTPV). AIP Conf. Proc., vol. 890, American Institute of PhysicsAIP; 2007, p. 297–316. https://doi.org/10.1063/1.2711748.

[18] Zhao B, Chen K, Buddhiraju S, Bhatt G, Lipson M, Fan S. High-performance near-field thermophotovoltaics for waste heat recovery. Nano Energy 2017;41:344–50. https://doi.org/10.1016/j.nanoen.2017.09.054.

[19] Bau S, Chen YB, Zhang ZM. Microscale radiation in thermophotovoltaic devices - A review. Int J Energy Res 2007;31:689–716. https://doi.org/10.1002/er.1286.

[20] Pan JL, Choy HKH, Fonstad CG. Very large radiative transfer over small distances from a black body for thermophotovoltaic applications. IEEE Trans Electron Devices 2000;47:241–9. https://doi.org/10.1109/16.817591.

[21] Lu Q, Zhou X, Krysa A, Marshall A, Carrington P, Tan CH, et al. InAs thermophotovoltaic cells with high quantum efficiency for waste heat recovery applications below 1000 °C. Sol Energy Mater Sol Cells 2018;179:334–8. https://doi.org/10.1016/j.solmat.2017.12.031.

[22] Vaillon R, Pérez J-P, Lucchesi C, Cakiroglu D, Chapuis P-O, Taliercio T, et al. Micron-sized liquid nitrogen-cooled indium antimonide photovoltaic cell for near-field

thermophotovoltaics. Opt Express 2019;27:A11. https://doi.org/10.1364/oe.27.000a11.

[23] Fiorino A, Zhu L, Thompson D, Mittapally R, Reddy P, Meyhofer E. Nanogap near-field thermophotovoltaics. Nat Nanotechnol 2018;13:806–11. https://doi.org/10.1038/s41565-018-0172-5.

[24] Inoue T, Koyama T, Kang DD, Ikeda K, Asano T, Noda S. One-Chip Near-Field Thermophotovoltaic Device Integrating a Thin-Film Thermal Emitter and Photovoltaic Cell. Nano Lett 2019;19:3948–52. https://doi.org/10.1021/acs.nanolett.9b01234.

[25] Bhatt GR, Zhao B, Roberts S, Datta I, Mohanty A, Lin T, et al. Integrated near-field thermo-photovoltaics for heat recycling. Nat Commun 2020;11:1–7. https://doi.org/10.1038/s41467-020-16197-6.

[26] Lucchesi C, Cakiroglu D, Perez J-P, Taliercio T, Tournié E, Chapuis P-O, et al. Near-Field Thermophotovoltaic Conversion with High Electrical Power Density and Cell Efficiency above 14%. Nano Lett 2021. https://doi.org/10.1021/acs.nanolett.0c04847.

[27] Fan D, Burger T, McSherry S, Lee B, Lenert A, Forrest SR. Near-perfect photon utilization in an air-bridge thermophotovoltaic cell. Nature 2020;586:237–41. https://doi.org/10.1038/s41586-020-2717-7.

[28] Wernsman B, Siergiej RR, Link SD, Mahorter RG, Palmisiano MN, Wehrer RJ, et al. Greater than 20% radiant heat conversion efficiency of a thermophotovoltaic radiator/module system using reflective spectral control. IEEE Trans Electron Devices 2004;51:512–5. https://doi.org/10.1109/TED.2003.823247.

[29] Omair Z, Scranton G, Pazos-Outón LM, Xiao TP, Steiner MA, Ganapati V, et al. Ultraefficient thermophotovoltaic power conversion by band-edge spectral filtering. Proc Natl Acad Sci U S A 2019;116:15356–61. https://doi.org/10.1073/pnas.1903001116.

[30] Xiao G, Zheng G, Qiu M, Li Q, Li D, Ni M. Thermionic energy conversion for concentrating solar power. Appl Energy 2017;208:1318–42. https://doi.org/10.1016/j.apenergy.2017.09.021.

[31] Xiao G, Zheng G, Ni D, Li Q, Qiu M, Ni M. Thermodynamic assessment of solar photon-enhanced thermionic conversion. Appl Energy 2018;223:134–45.

https://doi.org/10.1016/j.apenergy.2018.04.044.

[32] Ghashami M, Kwon S, Park K. Journal of Quantitative Spectroscopy & Radiative Transfer Near-field enhanced thermionic energy conversion for renewable energy recycling 2017;198:59–67. https://doi.org/10.1016/j.jqsrt.2017.04.033.

[33] Rahman E, Nojeh A. Semiconductor thermionics for next generation solar cells: photon enhanced or pure thermionic? Nat Commun 2021 121 2021;12:1–9. https://doi.org/10.1038/s41467-021-24891-2.

[34] Schwede JW, Bargatin I, Riley DC, Hardin BE, Rosenthal SJ, Sun Y, et al. Photon-enhanced thermionic emission for solar concentrator systems. Nat Mater 2010;9:762–7. https://doi.org/10.1038/nmat2814.

[35] Segev G, Rosenwaks Y, Kribus A. Limit of efficiency for photon-enhanced thermionic emission vs. photovoltaic and thermal conversion. Sol Energy Mater Sol Cells 2015;140:464–76. https://doi.org/10.1016/j.solmat.2015.05.001.

[36] Segev G, Rosenwaks Y, Kribus A. Efficiency of photon enhanced thermionic emission solar converters. Sol Energy Mater Sol Cells 2012;107:125–30. https://doi.org/10.1016/j.solmat.2012.08.006.

[37] Kribus A, Segev G. Solar energy conversion with photon-enhanced thermionic emission. J Opt (United Kingdom) 2016;18. https://doi.org/10.1088/2040-8978/18/7/073001.

[38] Vaillon R, Robin L, Muresan C, Ménézo C. Modeling of coupled spectral radiation, thermal and carrier transport in a silicon photovoltaic cell. Int J Heat Mass Transf 2006;49:4454–68. https://doi.org/10.1016/j.ijheatmasstransfer.2006.05.014.

[39] Park K, Basu S, King WP, Zhang ZM. Performance analysis of near-field thermophotovoltaic devices considering absorption distribution. J Quant Spectrosc Radiat Transf 2008;109:305–16. https://doi.org/10.1016/j.jqsrt.2007.08.022.

[40] Francoeur M, Pinar Mengüç M. Role of fluctuational electrodynamics in near-field radiative heat transfer. J Quant Spectrosc Radiat Transf 2008;109:280–93. https://doi.org/10.1016/j.jqsrt.2007.08.017.

[41]  Basu S, Zhang ZM, Fu CJ. Review of near-field thermal radiation and its application to energy conversion. Int J Energy Res 2009;33:1203–32. https://doi.org/10.1002/er.1607.

[42]  Francoeur M, Pinar Mengüç M, Vaillon R. Solution of near-field thermal radiation in one-dimensional layered media using dyadic Green's functions and the scattering matrix method. J Quant Spectrosc Radiat Transf 2009;110:2002–18. https://doi.org/10.1016/j.jqsrt.2009.05.010.

[43]  Francoeur M, Vaillon R, Meng MP. Thermal impacts on the performance of nanoscale-gap thermophotovoltaic power generators. IEEE Trans Energy Convers 2011;26:686–98. https://doi.org/10.1109/TEC.2011.2118212.

[44]  Varpula A, Prunnila M. Diffusion-emission theory of photon enhanced thermionic emission solar energy harvesters. J Appl Phys 2012;112:1–5. https://doi.org/10.1063/1.4747905.

[45]  Varpula A, Tappura K, Prunnila M. Si, GaAs, and InP as cathode materials for photon-enhanced thermionic emission solar cells. Sol Energy Mater Sol Cells 2015;134:351–8. https://doi.org/10.1016/j.solmat.2014.12.021.

[46]  Sze SM. Semiconductor Devices: Physics and Technology. second. John Wiley & Sons, New York; 2002.

[47]  Sze SM, Ng KK. Physics of Semiconductor Devices. 2nd ed. Hoboken, NJ, USA: John Wiley & Sons, Inc.; 2006. https://doi.org/10.1002/0470068329.

[48]  Hatsopoulos GN, Gyftopoulos EP. Thermionic Energy Conversion. Vol. 2: Theory, Technology, and Application. MIT Press; 1979.

[49]  Hatsopoulos GN, Gyftopoulos EP. Thermionic Energy Conversion. Vol. 1: Processes and Devices. MIT Press; 1973.

[50]  Su S, Wang Y, Liu T, Su G, Chen J. Space charge effects on the maximum efficiency and parametric design of a photon-enhanced thermionic solar cell. Sol Energy Mater Sol Cells 2014;121:137–43. https://doi.org/10.1016/j.solmat.2013.11.008.

[51]  Khoshaman AH, Koch AT, Chang M, Fan HDE, Moghaddam MV, Nojeh A.

Nanostructured Thermionics for Conversion of Light to Electricity: Simultaneous Extraction of Device Parameters. IEEE Trans Nanotechnol 2015. https://doi.org/10.1109/TNANO.2015.2426149.

[52] Lough BC, Lewis RA, Zhang C. Principles of charge and heat transport in thermionic devices. In: Al-Sarawi SF, editor. Smart Struct. Devices, Syst. II, vol. 5649, SPIE; 2005, p. 332. https://doi.org/10.1117/12.582101.

[53] Rahman E, Nojeh A. Interplay between Near-Field Radiative Coupling and Space-Charge Effects in a Microgap Thermionic Energy Converter under Fixed Heat Input. Phys Rev Appl 2020;14:024082. https://doi.org/10.1103/PhysRevApplied.14.024082.

[54] Laroche M, Carminati R, Greffet JJ. Near-field thermophotovoltaic energy conversion. J Appl Phys 2006;100:063704. https://doi.org/10.1063/1.2234560.

[55] Lim M, Jin S, Lee SS, Lee BJ. Graphene-assisted Si-InSb thermophotovoltaic system for low temperature applications. Opt Express 2015;23:A240. https://doi.org/10.1364/oe.23.00a240.

[56] Messina R, Ben-Abdallah P. Graphene-based photovoltaic cells for near-field thermal energy conversion. Sci Rep 2013;3:1–5. https://doi.org/10.1038/srep01383.

[57] Fu CJ, Tan WC. Near-field radiative heat transfer between two plane surfaces with one having a dielectric coating. J Quant Spectrosc Radiat Transf 2009;110:1027–36. https://doi.org/10.1016/j.jqsrt.2009.02.007.

[58] Narayanaswamy A, Chen G. Surface modes for near field thermophotovoltaics. Appl Phys Lett 2003;82:3544–6. https://doi.org/10.1063/1.1575936.

[59] Levinshtein ME, Rumyantsev SL, Shur M. Handbook Series On Semiconductor Parameters. vol. 1. Singapore: World Scientific Publishing Co. Pte. Ltd.; 1996.

[60] Bendelala F, Cheknane A, Hilal H. Enhanced low-gap thermophotovoltaic cell efficiency for a wide temperature range based on a selective meta-material emitter. Sol Energy 2018;174:1053–7. https://doi.org/10.1016/j.solener.2018.10.006.

[61] Khvostikov VP, Lunin LS, Kuznetsov V V., Ratushny VI, Oliva V, Khvostikova OA, et

al. InAs Based Multicomponent Solid Solutions for Thermophotovoltaic Converters. Tech Phys Lett 2003;29:851–2. https://doi.org/10.1134/1.1623867.

[62] Krier A, Yin M, Marshall ARJ, Kesaria M, Krier SE, McDougall S, et al. Low bandgap mid-infrared thermophotovoltaic arrays based on InAs. Infrared Phys Technol 2015;73:126–9. https://doi.org/10.1016/j.infrared.2015.09.011.

[63] Liao T, Zhang X, Chen X, Chen J. Near-field thermionic-thermophotovoltaic energy converters. J Appl Phys 2019;125. https://doi.org/10.1063/1.5086778.

[64] Chen K, Santhanam P, Sandhu S, Zhu L, Fan S. Heat-flux control and solid-state cooling by regulating chemical potential of photons in near-field electromagnetic heat transfer. Phys Rev B - Condens Matter Mater Phys 2015;91:1–8. https://doi.org/10.1103/PhysRevB.91.134301.

[65] Krishnamurthy S, Berding MA. Full-band-structure calculation of Shockley-Read-Hall recombination rates in InAs. J Appl Phys 2001;90:848–51. https://doi.org/10.1063/1.1381051.

[66] Sotoodeh M, Khalid AH, Rezazadeh AA. Empirical low-field mobility model for III-V compounds applicable in device simulation codes. J Appl Phys 2000;87:2890–900. https://doi.org/10.1063/1.372274.

[67] Lee JH, Bargatin I, Melosh NA, Howe RT. Optimal emitter-collector gap for thermionic energy converters. Appl Phys Lett 2012. https://doi.org/10.1063/1.4707379.

[68] Lee BJ, Zhang ZM. Lateral shifts in near-field thermal radiation with surface phonon polaritons. Nanoscale Microscale Thermophys Eng 2008;12:238–50. https://doi.org/10.1080/15567260802247505.

[69] Incropera FP, DeWitt DP. Fundamentals of heat and mass transfer. New York John Wiley & Sons; 2002.

[70] Concentrated solar power had a global total installed capacity of 6,451 MW in 2019 | REVE News of the wind sector in Spain and in the world n.d. https://www.evwind.es/2020/02/02/concentrated-solar-power-had-a-global-total-installed-capacity-of-6451-mw-in-2019/73360 (accessed April 27, 2021).

[71] Papapetrou M, Kosmadakis G, Cipollina A, La Commare U, Micale G. Industrial waste heat: Estimation of the technically available resource in the EU per industrial sector, temperature level and country. Appl Therm Eng 2018;138:207–16. https://doi.org/10.1016/j.applthermaleng.2018.04.043.

[72] Nicaise SM, Lin C, Azadi M, Bozorg-Grayeli T, Adebayo-Ige P, Lilley DE, et al. Micron-gap spacers with ultrahigh thermal resistance and mechanical robustness for direct energy conversion. Microsystems Nanoeng 2019;5. https://doi.org/10.1038/s41378-019-0071-4.

[73] Koeck FAM, Nemanich RJ, Lazea A, Haenen K. Thermionic electron emission from low work-function phosphorus doped diamond films. Diam Relat Mater 2009;18:789–91. https://doi.org/10.1016/j.diamond.2009.01.024.

[74] Datas A, Vaillon R. Thermionic-enhanced near-field thermophotovoltaics. Nano Energy 2019;61:10–7. https://doi.org/10.1016/j.nanoen.2019.04.039.

[75] Datas A, Vaillon R. Thermionic-enhanced near-field thermophotovoltaics for medium-grade heat sources. Appl Phys Lett 2019;114. https://doi.org/10.1063/1.5078602.